\documentclass[a4paper]{aa}

\begin{document} 
\thesaurus{06(08.02.1; 08.04.1; 08.05.02; 08.11.1; 13.25.5)}

\title{Hipparcos results on massive X-ray binaries}

\author{Claude Chevalier and Sergio A. Ilovaisky}
\institute{Observatoire de Haute-Provence (CNRS), F-04870 St.Michel
l'Observatoire, France\\
email: chevalier@obs-hp.fr}

\offprints{C.Chevalier}
\date{Received 28 August 1997, accepted 17 September 1997 }

\maketitle 

\begin{abstract}
We present results on parallaxes, magnitudes and proper motions for 17 optically identified massive X-ray binaries (MXRB) which were observed during the Hipparcos astrometry mission. This sample includes the sources detected in the medium-energy range (2-10 keV) having optical counterparts brighter than $V$=12. We compare the Hipparcos results with ground-based optical data and derive probable values for absolute magnitudes and peculiar tangential velocities ($v_{\mathrm{t}}$).  The 4 OB supergiant systems in our sample are high-velocity objects (the average value of $v_{\mathrm{t}}$ is greater than 60 km/s),  while the 13 Be systems have low velocities ($<$$v_{\mathrm{t}}$$>$ = 11.3 $\pm$ 6.7 km/s), suggesting a different formation mechanism for the two subgroups. The unusual X-ray/radio source LSI+61$^\circ$303 lies much closer than previously believed and its low optical luminosity ($M_{\mathrm{V}}$ = +2.2) and blue intrinsic colors now suggest it may be a peculiar type of low-mass X-ray binary.

\end{abstract}

\section{Introduction} 

In 1982 we proposed Hipparcos observations of galactic X-ray sources emitting in the 2-10 keV band and having a confirmed optical counterpart brighter than $V$=12. The Hipparcos Input Catalog (HIC) included 8 MXRB from our list. Since 1982 the number of optically identified MXRB increased considerably (see, for instance Tuohy et al. 1988). A cross-correlation between the catalogue of van Paradijs (1995) and the HIC reveals that 9 additional X-ray sources (all Be systems) were observed as survey B stars during the Hipparcos mission. Among the 17 O or B stars in the HIC which are up to now optically identified with MXRB, 13 are classified as Be stars (the Be subgroup) and 4 as OB supergiants (the SG subgroup).

\begin{table*}[t]
\renewcommand{\footnoterule}{\rule{5mm}{0mm}\vspace{-2mm}}
\caption[]{Supergiant X-ray binaries : Ground-based distance estimates. Hipparcos  proper motions. Peculiar tangential velocities} 
\begin{flushleft}
\begin{minipage}{18cm}
\vspace{-\abovedisplayskip}
\begin{tabular}[t]{cllrrrlll}
\\[-2mm]
\hline
\\[-3mm]
X-ray & Name & $d_{\mathrm{est}}\pm\sigma$ & $\pi_{\mathrm{H}} \pm\sigma$ & $\mu_{\mathrm{\alpha}}\cos\delta \pm\sigma$ & $\mu_{\mathrm{\delta}}\pm\sigma$ & $v_{\mathrm{t,\alpha}}\pm\sigma$ & $v_{\mathrm{t,\delta}}\pm\sigma$ & $v_{\mathrm{t}}\pm\sigma$   \\
source &  & (kpc) & (mas) & (mas/y) & (mas/y) & (km/s) & (km/s) & (km/s)  \\
\hline
\\[-3mm]
0114+650   & V662 Cas  & $^{1.4~0.5\footnote{van Oijen 1989, Bradt \& McClintock 1983}}_{7.0~1.0\footnote{Reig et al. 1996}}$   &  0.88 ~1.74   & $-$1.53 ~1.59 & +1.58 ~1.28   & $^{-21.4 ~10.8}_{-44.5~53.0}$ & $^{+19.4~9.4}_{+62.9~43.0}$ & $^{28.9~10.2}_{77.0~47.0}$ \\[+2mm]
           &           & 3.8\footnote{distance corresponding to $M_{\mathrm{V}}$ $\sim$ $-$6.2}                &               &               &               & $-$31.3                 & +38.1                  & 49.3 \\[+1mm]
0900$-$403 & GP Vel    & 1.4 ~0.5                 & $-$0.38 ~0.78 & $-$5.81 ~0.58 & +8.25 ~0.66   & $-$22.6 ~12.7 & +59.9 ~22.0   & 64.0 ~21.1 \\
1700$-$377 & V884 Sco  & 1.7 ~0.5                 & $-$0.21 ~0.86 & +1.90 ~0.81   & +4.71 ~0.53   & +19.0 ~7.8    & +54.1 ~11.7   & 57.3 ~11.3 \\
1956+350   & V1357 Cyg & 2.5 ~0.5                 &  0.58 ~1.01   & $-$3.82 ~0.79 & $-$7.62 ~0.91 & $-$47.0 ~12.3 & $-$80.7 ~19.3 & 93.4 ~17.8 \\
\hline
\end{tabular}
\end{minipage}
\end{flushleft}
\end{table*}

\begin{table*}[t]
\renewcommand{\footnoterule}{\rule{5mm}{0mm}\vspace{-2mm}}
\caption[]{Supergiant X-ray binaries : Ground-based magnitudes, colors and spectra. Hipparcos magnitudes and visual absolute magnitudes} 
\begin{flushleft}
\begin{minipage}{18cm}
\vspace{-\abovedisplayskip}
\begin{tabular}[t]{cllllllrrll}
\\[-2mm]
\hline
\\[-3mm]
 X-ray  & $V$   & $B$$-$$V$ & $U$$-$$B$ & $E$($B$$-$$V$) & $A_{\mathrm{V}}$ & Spectral & $V_{\mathrm{H,med}}$ & $\Delta V_{\mathrm{H}}$ & $d_{\mathrm{est}}\pm\sigma$ & $M_{\mathrm{V}}\pm\sigma$  \\
source & (mag) & (mag)     & (mag)     & (mag)        & (mag)   &  Type       & (mag)       & (mag)          & (kpc)              & (mag) \\
\hline
\\[-3mm]
0114+650    & 11.0 & 1.2  & +0.1    & 1.4  & 4.3  & B0.5 Ib  & 11.05 & 0.15     & $^{1.4~0.5}_{7.0~1.0}$ & $^{-4.0~0.9}_{-7.5~0.6}$ \\[+0.5mm]
            &      &      &         &      &      &          &       &          & 3.8\footnote{distance corresponding to $M_{\mathrm{V}}$ $\sim$ $-$6.2}              & $-$6.2\footnote{assumed}     \\
0900$-$403  & 6.9  & 0.47 & $-$0.51 & 0.7  & 2.2  & B0.5 Ib  & 6.91  & 0.12     & 1.4 ~0.5               & $-$6.0 ~0.8  \\
1700$-$377  & 6.6  & 0.27 & $-$0.72 & 0.52 & 1.6  & O6.5f    & 6.48  & 0.07     & 1.7 ~0.5               & $-$6.2 ~0.7  \\
1956+350    & 8.9  & 0.84 & $-$0.26 & 1.06 & 3.3  & O9.7 Iab & 8.84  & 0.09     & 2.5 ~0.5               & $-$6.4 ~0.5   \\
\hline
\end{tabular}
\end{minipage}
\end{flushleft}
\end{table*}

\section{Hipparcos results}
The parallaxes $\pi_{\mathrm{H}}$ (and corresponding distances $d_{\mathrm{H}}$) and the proper motions ($\mu_{\mathrm{\alpha}}\cos\delta$ and $\mu_{\mathrm{\delta}}$)  obtained with the Hipparcos satellite (Hipparcos Catalogue 1997) are given in Tables 1 and 3 for the 17 objects in our sample. Tables 2 and 4 summarize the ground-based magnitudes, colors and spectra for the two subgroups, taken essentially from the compilation by van Paradijs (1995). Numbers shown in italics in Tables 3 and 4 correspond to the two Be systems whose parallaxes are not significant and for which distances were estimated as explained below. Colons following certain entries in Table 4 indicate uncertain values. No significant parallaxes were measured for any of the members of the SG subgroup.

Tables 1 and 3 give the peculiar tangential velocities,  $v_{\mathrm{t,\alpha}}$, $v_{\mathrm{t,\delta}}$ and $v_{\mathrm{t}}$,
computed from the observed proper motions. They have been corrected for differential galactic rotation and solar motion (see Green 1985 and van de Kamp 1967). For the SG subgroup previously published distances have been assumed (see below).

Hipparcos $V$ magnitudes in the Johnson system (median values) and the range of variability  ($\Delta V_{\mathrm{H}} = V_{\mathrm{min}}-V_{\mathrm{max}}$) recorded during the mission are given in Tables 2 and 4. The median values were used to derive new absolute visual magnitudes assuming the new distances and allowing for the visual interstellar extinction given in the Tables.

Quiescent X-ray luminosities for the members of the Be subgroup given in Table 4 and corresponding to the new distances were computed using the 2-10 keV fluxes given in the compilation of van Paradijs (1995). The values given in parenthesis refer to the outburst state.

\section{Discussion of the supergiant (SG) subgroup}

\subsection{Distances and absolute magnitudes}

The four stars in this subgroup all have Hipparcos parallaxes which are not significant since their values are either negative or smaller than the stated uncertainties (see Table 1), indicating a distance probably larger than 1 kpc to these stars. There is a general agreement on the supergiant nature of GP Vel (Vela X-1), V884 Sco (1700$-$377) and V1357 Cyg (Cygnus X-1). The distance estimates given by Bradt and McClintock (1983) or van Oijen (1989) correspond to absolute visual magnitudes between $-$6.0 and $-$6.4 for the optical SG counterparts.

The case of V662 Cas (0114+65) is less clear since it shares properties with the two subgroups (SG and Be). Assuming $M_{\mathrm{V}}$ $\sim$ $-$6.2 for V662 Cas (the average for the other three SG systems) yields $d \sim$ 3.8 kpc, a distance intermediate between the estimates of Bradt and McClintock and van Oijen, 1.4 $\pm$ 0.5 kpc, and that of Reig et al. (1996), 7 $\pm$ 1 kpc, which appears to be an overestimate.

\subsection{Peculiar tangential velocities of the SG subgroup}

From the proper motions measured with Hipparcos we have computed the peculiar tangential velocities of the SG stars after subtraction of the terms due to differential galactic rotation and to the solar velocity toward the apex. The results are listed in Table 1. With the currently adopted distance estimates, GP Vel, V884 Sco and V1357 Cyg have peculiar tangential velocities larger than 50 km/s. The value for V662 Cas is about 49 km/s for a distance of $d \sim$ 3.8 kpc. This shows that the four X-ray binaries in the SG subgroup have high peculiar tangential velocities (average value of $v_{\mathrm{t}}$ for this subgroup is 66 $\pm$ 19 km/s).

This result is much more reliable than previous studies based on radial velocities, since the atmospheric layers in which the optical absorption lines are formed have variable outflow velocities which can easily reach 30 km/s (see van Oijen 1989). This result appears to support the assumption that the SG X-ray binaries come from massive OB binaries which have survived a supernova explosion, in agreement with expectations
from binary evolution (van den Heuvel 1994).

The proper motion components measured with Hipparcos for GP Vel are quite different from previous ground-based estimates (see Table 1 in Kaper et al. 1997) and do not support the suggestion by Kaper et al. that it might originate from the OB association Vela OB1.

\section{Discussion of the Be subgroup}

\subsection{Distances and absolute magnitudes}

\subsubsection{Sources with distances agreeing with some previous estimates}

Eight sources, $\gamma$ Cas (0053+604), X Per (0352+309), HD 63666 (0739$-$529), HD 56553 (0749$-$600), HD 91188 \linebreak (1036$-$565), BZ Cru (1249$-$637), HD 109857 (1253$-$761) and $\mu^{2}$ Cru  (1255$-$567), have Hipparcos distances ($d_{\mathrm{H}}$) (see Table 3) in agreement with (at least) some of the estimates previously derived from the spectral types and reddening obtained from ground-based and IUE data.

\begin{table*}[t]
\renewcommand{\footnoterule}{\rule{5mm}{0mm}\vspace{-2mm}}
\caption[]{Be X-ray binaries : Hipparcos parallaxes and proper motions.
Derived distances and peculiar tangential velocities} 
\begin{flushleft}
\begin{minipage}{18cm}
\vspace{-\abovedisplayskip}
\begin{tabular}[t]{clrrrrrrrl}
\\[-2mm]
\hline
\\[-3mm]
X-ray & Name & $\pi_{\mathrm{H}} \pm\sigma$ & $\mu_{\mathrm{\alpha}}\cos\delta \pm\sigma$ & $\mu_{\mathrm{\delta}}\pm\sigma$ & $v_{\mathrm{t,\alpha}}\pm\sigma$ & $v_{\mathrm{t,\delta}}\pm\sigma$ & $v_{\mathrm{t}}\pm\sigma$ & $d_{\mathrm{H}}$ & $^{\mathrm{max}}_{\mathrm{min}}$  \\
source& & (mas) & (mas/y) & (mas/y) & (km/s) & (km/s) & (km/s) & (pc) & (pc) \\
\hline
\\[-3mm]
0053+604 & $\gamma$ Cas & 5.32 ~0.56      & 25.65 ~0.42     & $-$3.82 ~0.44  & 7.48 ~2.64     & 4.28 ~0.67    & 8.6 ~2.3   & 188      & $^{208}_{168}$   \\[+0.5mm]
0236+610 & LSI+61$^\circ$303 & 5.65 ~2.28 & 0.62 ~1.95      &    1.63 ~1.75  & $-$12.03 ~1.80 & 15.00 ~1.73   & 19.2 ~1.8  & 177      & $^{300}_{130}$   \\[+0.5mm]
0352+309 & X Per & 1.21 ~0.94              & $-$1.99 ~0.91  & $-$4.51 ~0.88  & $-$17.33 ~7.44 & $-$2.34 ~14.2 & 17.5 ~7.6 & 830       & $^{3700}_{500}$   \\[+0.5mm]
0521+373 & HD 34921 & {\it0.39 ~0.89}\footnote[1]{not significant}      & 0.02 ~ 1.07    & $-$5.21 ~0.49  & {\it $-$3.7 ~~~~~~~} & {\it $-$8.0 ~~~~~~~} & {\it 8.8 ~~~~~}& {\it1050}\footnote[2]{assumed distance} & \\[+0.5mm]
0535+262 & V725 Tau & 3.00 ~1.72           & $-$5.08 ~1.65  & $-$4.41 ~1.02  & $-$10.29 ~5.39 & 8.56 ~4.56    & 13.4 ~5.1 & 330    & $^{780}_{210}$    \\[+0.5mm]
0739$-$529 & HD 63666 & 1.94 ~0.55         & $-$5.63 ~0.61  & 9.86 ~0.60     & $-$5.63 ~3.86  & 20.25 ~7.63   & 21.0 ~7.4& 520     & $^{720}_{400}$     \\[+0.5mm]
0749$-$600 & HD 65663 & 2.49 ~0.50         & $-$5.14 ~0.51  & 13.69 ~0.49    & $-$1.46 ~2.07  & 19.79 ~5.69   & 19.8 ~5.7 & 400    & $^{500}_{330}$    \\[+0.5mm]
1036$-$565 & HD 91188 & 1.01 ~0.57         & $-$5.60 ~0.56  & $-$0.63 ~0.49  & $-$7.19 ~12.8  & $-$0.90 ~2.31 & 7.2 ~12.8& 1000    & $^{2300}_{600}$     \\[+0.5mm]
1145$-$619 & V801 Cen & 1.98 ~0.95         & $-$6.47 ~0.82  & 2.06 ~0.83     & 2.83 ~6.81     & 8.66 ~3.28  & 9.1 ~3.8 & 500       & $^{1000}_{300}$    \\[+0.5mm]
1249$-$637 & BZ Cru    & 3.32 ~0.56        & $-$13.32 ~0.39 & $-$3.47 ~0.46  & $-$1.78 ~3.20  & 1.67 ~1.08  & 2.4 ~2.5  & 300      & $^{360}_{260}$     \\[+0.5mm]
1253$-$761 & HD 109857 & 4.24 ~0.60        & $-$27.07 ~0.66 & $-$10.01 ~0.67 & $-$13.25 ~4.32 & 6.41 ~1.74  & 14.7 ~4.0& 236       & $^{275}_{207}$      \\[+0.5mm]
1255$-$567 & $\mu^{2}$ Cru & 9.03 ~0.61    & $-$32.35 ~0.40 & $-$10.93 ~0.41 & $-$0.58 ~1.42  & 2.25 ~0.61  &  3.3 ~0.7 & 111      & $^{119}_{104}$     \\[+0.5mm]
2202+501 & SAO 51568 & $-${\it0.24 ~0.84}$^{a}$  & 2.32 ~0.75     & $-$0.30 ~0.72  & {\it$-$3.1 ~~~~~~~} & {\it$-$1.7 ~~~~~~~} & {\it 3.5 ~~~~~}& {\it700}$^{b}$ &   \\[+0.5mm]
\hline
\end{tabular}
\end{minipage}
\end{flushleft}
\end{table*}

\begin{table*}[t]
\renewcommand{\footnoterule}{\rule{5mm}{0mm}\vspace{-2mm}}
\caption[]{Be X-ray binaries : Ground-based magnitudes, colors and spectra. 
Hipparcos magnitudes. New visual absolute magnitudes and X-ray luminosities} 
\begin{flushleft}
\begin{minipage}{18cm}
\vspace{-\abovedisplayskip}
\begin{tabular}[t]{clrrlllrrll}
\\[-2mm]
\hline
\\[-3mm]
X-ray & $V$ & $B$$-$$V$ & $U$$-$$B$ & $E$($B$$-$$V$) & $A_{\mathrm{V}}$ & Spectral & $V_{\mathrm{H,med}}$ & $\Delta V_{\mathrm{H}}$ &  $M_{\mathrm{V}} \pm\sigma$ & $\log L_{\mathrm{x(2-10 keV)}}$ \\
source & (mag) & (mag) & (mag) & (mag) & (mag) & Type & (mag) & (mag) & (mag) & (erg/s) \\
\hline
\\[-3mm]
0053+604   & 1.6-3.0 & $-$0.15 & $-$1.08 & 0.05 & 0.16 & B0.5e III-IV & 2.15  & 0.04  & $-$4.4 ~0.3      &  32.66 (33.00)  \\
0236+610   & 10.7    & +0.80     & $-$0.30  & 0.75 & 2.3  & B0e:         & 10.70 & 0.17  & +2.2 ~1.3        &  31.20        \\
0352+309   & 6.0-6.6 & +0.29    & $-$0.82 & 0.40  & 1.3  & O9e III-IV   & 6.79  & 0.07  & $-4$.0 ~1.7      &  34.20 (34.82)  \\
0521+373   & 7.51    & +0.14    & $-$0.86 & 0.42 & 1.3  & B0pe IV      & 7.40  & 0.12  & ${\it-4.0}\footnote[1]{from assumed distance}$ &  {\it 33.46}  \\
0535+262   & 8.9-9.6 & 0.45-0.62 & $-$0.54 & 0.75 & 2.3  & O9.7e II   & 9.20  & 0.22  & $-$0.7 ~1.3      &  32.93 (35.90)  \\
0739$-$529 & 7.62    & +0.02    & $-$0.24 & 0.10:  & 0.3: & B7e IV-V   & 7.60  & 0.04  & $-$1.3 ~0.7      &  32.69        \\
0749$-$600 & 6.73    & +0.05    & $-$0.25 & 0.09   & 0.3  & B8e III    & 6.73  & 0.02  & $-$1.6 ~0.5      &  32.46        \\
1036$-$565 & 6.64    & $-$0.10 & $-$0.56 & 0.05:  & 0.16 & B4e III    & 6.62  & 0.12  & $-$3.5 ~1.2      &  33.93        \\
1145$-$619 & 9.30    & +0.18    & $-$0.81 & 0.35   & 1.1  & B1e V      & 8.86  & 0.25  & $-$0.7 ~1.2      &  33.41 (35.81)  \\
1249$-$637 & 5.31    & +0.27    & $-$0.79 & 0.40   & 1.2  & B0e III    & 5.27  & 0.05  & $-$3.3 ~0.5      &  32.71        \\
1253$-$761 & 6.49    & +0.08    & $-$0.24 & 0.16   & 0.5  & B7e V      & 6.46  & 0.03  & $-$0.9 ~0.4      &  31.94        \\
1255$-$567 & 5.17    & $-$0.12 & 0.51    & 0.04   & 0.12 & B5e V      & 5.08  & 0.03  & $-$0.3 ~0.2      &  31.41        \\
2202+501   & 8.80    & +0.167\footnote[2]{from the Tycho Catalog} &      & 0.30:   &1.0:  & B5:\footnote[3]{from Simbad database}        & 9.27  & 0.12  & {\it$-$1.0}$^{a}$ &  {\it 32.95}  \\
\hline
\end{tabular}
\end{minipage}
\end{flushleft}
\end{table*}

For $\gamma$ Cas, $d_{\mathrm{H}}$ = 188 $\pm$ 20 pc agrees with the oldest estimates of Schmidt-Kaler (1964), Lesh (1968), Bohlin (1970) and Hutchings (1970) but is smaller than the currently adopted value of 300 pc (Moffat et al. 1973, Bradt and McClintock 1983, van Oijen 1989). From the Hipparcos distance, $A_{\mathrm{V}} \sim$ 0.16 and $V_{\mathrm{H}}$ = 2.15 we derive $M_{\mathrm{V}} = -4.4 \pm 0.3$, a value consistent with a B0 III-IV spectral type (Gray 1992).

For X Per, $d_{\mathrm{H}}$ = 830 pc (with a large uncertainty) is in agreement with the estimate given by van Oijen (1989). Such a distance is intermediate between the value given by Brucato and Kristian (1972), $\sim$ 350 pc, close to the distance to the Perseus OB2 association --to which it does not belong-- and that derived by Fabregat et al. (1992), 1.3 $\pm$ 0.2 kpc, which may be an overestimate. From the Hipparcos distance we derive $M_{\mathrm{V}} \sim -4 \pm 1.7$, consistent with a spectral type in the range O8 to B2.

The distance to HD 91188, $d_{\mathrm{H}} \sim$ 1 kpc, yields $M_{\mathrm{V}} = -3.6 \pm 1.2$, consistent with an early B giant, while that for BZ Cru, $d_{\mathrm{H}}$ = 300 pc, yields $M_{\mathrm{V}} = -3.3 \pm 0.5$, which is not in contradiction with a B0 spectral type.

For HD 63666, $d_{\mathrm{H}}$ = 520 pc, yields $M_{\mathrm{V}}  = -1.3 \pm 0.7$, consistent with the spectral type of a mid-B dwarf, while the distance for HD 65663, $d_{\mathrm{H}}$ = 400 pc, yields $M_{\mathrm{V}} = -1.6 \pm 0.5$, which may be consistent with a late B giant.

The distance to HD 109857, $d_{\mathrm{H}}$ = 236 pc, yields $M_{\mathrm{V}} = -0.9 \pm 0.4$, consistent with a spectral type B5 to B7V and that for $\mu^{2}$ Cru, $d_{\mathrm{H}}$ = 111 pc, yields $M_{\mathrm{V}} = -0.3 \pm 0.2$, consistent with a B6-B8 V spectral type.

\subsubsection{Sources with distances smaller than previous estimates}

{\bf V725 Tau (0535+262)}

The Hipparcos parallax of V725 Tau yields a distance $d_{\mathrm{H}}$ = 330 pc (1$\sigma$ interval : 210-780 pc), much smaller than the previously adopted distances of 1.3 kpc (Hutchings et al. 1978), 1.8 kpc (Giangrande et al. 1980 and van Oijen 1989), 2.0 kpc (Giovannelli and Graziati 1992), and 2.4 $\pm$ 0.4 kpc (Reig 1996). The reddening of the star, $E$($B$$-$$V$) = 0.75 $\pm$ 0.05, was derived by Giovannelli et al. (1980) using the 2200 \AA\ interstellar band and the interstellar extinction law of Seaton (1979). Combining this reddening with the observed color of $B$$-$$V$ = $+$0.51 (0.45 to 0.62) gives an unreddened color index of ($B$$-$$V$)$_{0}$ = $-$0.24 ($-$0.30 to $-$0.13), corresponding to a B2 spectral type (B0 to B5), in fair agreement with the B0V spectral type found by Hutchings et al. from the absorption lines, but less negative than expected from an O9.5-7 II-III star (see Giovannelli \& Graziati 1992). Similarly, the absolute visual magnitude corresponding to a distance $\leq 1$ kpc derived from $\pi_{\mathrm{H}}$ is more consistent with an early B dwarf star than with a late O giant.
\\

{\bf V801 Cen (1145-619)}

The Hipparcos parallax of V801 Cen yields $d_{\mathrm{H}} \sim$  500 pc (1$\sigma$ interval : 300-1000 pc). Previous estimates were 1.5 kpc (Hammerschlag-Hensberge et al. 1980, Lamb et al. 1980), 2.1 kpc (van Oijen 1989, Bradt and McClintock 1983) and 3.1 $\pm$ 0.3 kpc (Stevens et al. 1997). Again, the Hipparcos parallax indicates a distance closer to 1 kpc than to 2-3 kpc.
\\

{\bf V615 Cas (LSI+61$^{\circ}$303)}

The case of V615 Cas is still more intriguing. Hipparcos observations ($\pi$ = 5.65 $\pm$ 2.28 mas) yield a distance of $d_{\mathrm{H}}$ = 177 pc (1$\sigma$ interval : 130-300 pc). Even a 2$\sigma$ error would give a distance less than 1 kpc. The small distance derived from the Hipparcos data is in strong disagreement with previous estimates. Frail and Hjellming (1991) proposed 2.0 $\pm$ 0.2 kpc on the basis of H I and C$^{18}$O emission and absorption in the spectra of the associated radio source. Gregory et al. (1979) proposed 2.3 kpc assuming the same distance to the source as for IC 1805. 

The reddening $E$($B$$-$$V$) = 0.75 $\pm$ 0.05 has been derived from IUE spectra by Howarth (1983) using the 2200 \AA\ bump. The absolute magnitude corresponding to the Hipparcos distance is then $M_{\mathrm{V}}$ = +2.2 $\pm$ 1.3, which is not consistent with the absolute magnitude of an early B star. The intrinsic colors ($B$$-$$V$)$_{0}$ = +0.05 and ($U$$-$$B$)$_{0} \sim -0.8$ do not correspond to any normal star but are typical of accretion disks around compact objects, which also have absolute visual magnitudes in the range 0 to +3 (van Paradijs and McClintock 1995).

An interstellar reddening $E$($B$$-$$V$) $\leq$ 0.4 is expected (Ishida 1969) if the system were closer than 1 kpc, unless part of it is of circumstellar origin, which is not excluded given the observed infrared excess (Elias et al. 1985, D'Amico et al. 1987, Paredes et al. 1994, Marti and Paredes 1995).

The optical spectrum (Gregory et al. 1979, Hutchings and Crampton 1981, Steele et al. 1995) ressembles that of a B1 Ib star with strong double-peaked emission at H$\alpha$ and H$\beta$. The wings of H$\alpha$ extend over 50 \AA\ (Gregory et al.) implying Doppler broadening velocities of 1100 km/s. The width of these emission lines is emphasized by Hutchings and Crampton, who remark that although no known Be star shows an emission line profile as broad as in LSI+61$^{\circ}$303, such profiles are common in cataclysmic variables in which a disk forms around a white dwarf. It is tempting to propose an alternative model for LSI+61$^{\circ}$303 where the primary would be a compact object surrounded by a thick accretion disk in rapid Keplerian rotation having an effective temperature of $T_{\mathrm{eff}}$ $\sim$ 15000 K and hiding the central X-ray source. The strong periodic and non-thermal radio emission seems to rule out a cataclysmic variable and exhibits similarities with the radio emission from binary X-ray stars such as SS~443 or Cir~X-1. The 26-d period could be related to disk precession.

\subsubsection{Sources with no significant parallax}

Two sources (values in italics in Table 3) have no significant Hipparcos parallax (both are smaller than the uncertainties of about 0.9 mas), 
HD~34921 (0521+373) and SAO~51568  (2202+501), indicating that the distances to these stars are larger than 500 pc. Assuming $M_{\mathrm{V}} \sim  -4$ for HD 34921, a B0 subgiant, yields a distance of 1.05 kpc. Similarly, assuming $M_{\mathrm{V}} \sim -1$ for SAO 51568 (of B5 spectral type) yields 700 pc. We adopt these distances to derive the peculiar tangential velocities of these stars from their Hipparcos proper motions.

\subsubsection{X-ray luminosities for the Be subgroup}

The distance determinations for the 13 Be systems given in Table 3 imply quiescent X-ray luminosities (Table 4) which are on the average lower ($<$$\log L_{\mathrm{x(2-10 keV)}}$$>$ = 32.77 $\pm$ 0.89) than previously published estimates (33.3 $\pm$ 1.2 in van Paradijs and McClintock 1995 --although for a larger sample--).

\subsection{Peculiar tangential velocities of the Be subgroup}

Peculiar tangential velocities of the stars in the Be subgroup (Table 3) have been computed from the proper motion components measured with Hipparcos after subtraction of the terms due to solar motion and to differential galactic rotation. Distances for all objects have been derived from Hipparcos parallaxes, with the exception of HD 34921 and SAO 51568 for which distances of 1050 pc and 700 pc have been adopted as described above. The quoted uncertainties result from the standard deviations on the parallaxes and the proper motion components. All the peculiar tangential velocities of the Be X-ray stars are found $\leq$ 21 km/s. The average peculiar tangential velocity for this sample of 13 stars is $<$$v_{\mathrm{t}}$$>$ = 11.3 km/s with a dispersion of 6.7 km/s. Such a result shows unambiguously that the Be X-ray stars are ``normal'' low-velocity stars of the solar neighborhood (d $\leq$ 1 kpc).

\section{Conclusions}

We distinguished two subgroups in the sample of 17 optically identified MXRB observed during the Hipparcos astrometric mission : (SG) 4 OB supergiants and (Be) 13 Be X-ray stars. The study of the astrometric data yields the following results :

Stars of the SG subgroup are luminous ($M_{\mathrm{V}} \leq -6$) OB supergiants at distances larger than 1 kpc (Hipparcos parallaxes are either negative or not significant). Assuming a distance consistent with the spectral type and reddening, the proper motion components yield peculiar tangential velocities larger than 50 km/s, confirming them to be high-velocity objects (The average value for the four systems is $<$$v_{\mathrm{t}}$$>$ = 66 $\pm$ 19 km/s). This result supports the assumption that supergiant MXRBs originate from massive OB binaries which have survived a supernova explosion (van den Heuvel 1994).

In contrast, stars of the Be subgroup are relatively nearby systems generally situated at distances $\leq$ 1 kpc. Their peculiar tangential velocities are much lower (average 11.3 $\pm$ 6.7 km/s). They appear as ``normal'' low-velocity stars of the solar neighborhood. This result does not favor the assumption that a supernova explosion has occurred in these systems but could indicate that many Be stars in this subgroup may  have a white dwarf as a companion and this would be consistent with their low X-ray luminosities (see Table 4, where X Per, which seems to harbor a neutron star, stands out as the brightest source). Such a suggestion has been made before (Waters et al. 1989), particularly in the case of $\gamma$ Cas (Murakami et al. 1986,  Haberl 1995).

One object, LSI+61$^{\circ}$303, is definitely singular and different from the others. Much nearer than expected ($<$ 1 kpc), its absolute visual magnitude, intrinsic colors and very broad, double-peaked Balmer emission lines are quite similar to those of accretion disks around collapsed stars. We thus suggest that LSI+61$^{\circ}$303 could be a low-mass X-ray binary system in which a thick accretion disk around the compact object mimics a B star, the system being surrounded by a cool envelope responsible for the shell spectrum and part of the IR excess and reddening.


\begin{thebibliography}{}


\bibitem{}Bohlin R.C. 1970, ApJ 162, 571
\bibitem{}Bradt H.V., McClintock J.E. 1983, Ann. Rev. Astr. Ap. 21, 13
\bibitem{}Brucato R.J., Kristian J. 1972, ApJL 173, L105
\bibitem{}D'Amico N. et al. 1987, A\&A 180, 114
\bibitem{}Elias J.H. et al. 1985, AJ 90, 1188
\bibitem{}Fabregat J. et al. 1992, A\&A 259, 522
\bibitem{}Frail D.A., Hjellming R.M. 1991, AJ 101, 2126
\bibitem{}Giangrande A. et al. 1980, A\&AS 40, 289
\bibitem{}Giovannelli F., Sabau Graziati L. 1992, Sp. Sci. Rev. 59, 1

\bibitem{}Giovannelli F. et al. 1980, in Proc. 2nd European IUE Conf., ESA SP-157, p.159 

\bibitem{}Gray D.F. 1992, The Observation and Analysis of Stellar Photospheres, Cambridge University Press, p.481

\bibitem{}Gregory P.C. et al. 1979, AJ 84, 1080
\bibitem{}Green R.M. 1985, Spherical Astronomy, Cambridge University Press
\bibitem{}Haberl F. 1995, A\&A 296, 685 
\bibitem{}Hammerschlag-Hensberge G. et al. 1980, A\&A 85, 119
\bibitem{}van den Heuvel E.P.J. 1994, in Interacting Binaries, S.N.Shore, M.Livio, E.P.J.van den Heuvel (Eds.), Springer-Verlag, p. 379
\bibitem{}Hipparcos Catalogue 1997, ESA SP-1200
\bibitem{}Howarth I.D. 1983, MNRAS 203, 801
\bibitem{}Hutchings J.B. 1970, MNRAS 150, 55
\bibitem{}Hutchings J.B., Crampton D. 1981, PASP 93, 486
\bibitem{}Hutchings J.B. et al. 1978, ApJ 223, 530
\bibitem{}Ishida K. 1969, MNRAS 144, 55
\bibitem{}van de Kamp P. 1967, Principles of Astrometry, Freeman Press, San Francisco
\bibitem{}Kaper L. et al. 1997 ApJL 475, L37
\bibitem{}Lamb R.C. 1980, ApJ 239, 651
\bibitem{}Lesh J.R. 1968, ApJS 17, 371
\bibitem{}Marti J., Paredes, J.M. 1995, A\&A 298, 151
\bibitem{}Moffat A.F.J., Haupt W., Schmidt-Kaler, T. 1973, A\&A 23, 433
\bibitem{}Murakami T. et al. 1986, ApJ 310, L31
\bibitem{}van Oijen J.G.J. 1989, A\&A 217, 115

\bibitem{}van Paradijs J. 1995, in X-Ray Binaries, W.H.G Lewin, J. van Paradijs, E.P.J. van den Heuvel (Eds.), Cambridge University Press, p.536

\bibitem{}van Paradijs J. and McClintock J.E. 1995, in X-Ray Binaries, W.H.G Lewin, J. van Paradijs, E.P.J. van den Heuvel (Eds.), Cambridge University Press, p.58

\bibitem{}Paredes J.M. et al. 1994, A\&A 288, 519
\bibitem{}Reig P. 1996, PASP 108, 639
\bibitem{}Reig P. et al. 1996, A\&A 311, 879
\bibitem{}Seaton M. 1979, MNRAS 187, 759
\bibitem{}Schmidt-Kaler T. 1964, Veroff. Univ. Sternw. Bonn, No.70 
\bibitem{}Steele I.A. et al. 1996, A\&AS 120, 213
\bibitem{}Stevens J.B. et al. 1997, submitted to MNRAS

\bibitem{}Tuohy, I.R. et al. 1988, in Physics of Neutron Stars and Black Holes, Universal Academy Press, Tokyo, p.93
\bibitem{}Waters I.B.F.M. et al. 1989, A\&A 220, L1

\end{thebibliography}
\end{document}